\documentclass[prb,twocolumn,showpacs,showkeys,superscriptaddress]{revtex4}
\usepackage{graphicx}
\usepackage{longtable}
\usepackage{abbreviations}
\begin{document}
%\preprint{number}

\title{EXAFS study of lead-free relaxor ferroelectric
BaTi$_{1-x}$Zr$_{x}$O$_{3}$ at the Zr K-edge}
\author{C.~Laulh\'e}
\author{F.~Hippert}
\author{J. Kreisel}
\email{jens.kreisel@inpg.fr} \affiliation{Laboratoire des
Mat\'eriaux et du G\'enie Physique (CNRS), ENSPG, B.P. 46, F-38402
Saint Martin d'H\`eres Cedex, France}
\author{M. Maglione}
\author{A. Simon}
\affiliation{Institut de Chimie de la Mati\`ere Condens\'ee de
Bordeaux (CNRS), 87 avenue A. Schweitzer, F-33608 Pessac, France}
\author{J.L. Hazemann}
\affiliation{Laboratoire de Cristallographie (CNRS), 25 rue des
Martyrs, BP 166, F-38042 Grenoble cedex 9, France}
\author{V. Nassif}
\affiliation{CEA Grenoble, DRFMC/SP2M/NRS, 17 rue des Martyrs,
F-38054 Grenoble cedex 9, France}
\date{\today}

\begin{abstract}

Extended X-ray absorption fine structure (EXAFS) experiments at
the Zr K-edge were carried out on perovskite relaxor
ferroelectrics BaTi$_{1-x}$Zr$_x$O$_3$ (BTZ) ($x$ = 0.25, 0.30,
0.35), and on BaZrO$_3$ for comparison. Structural information up
to 4.5 $\rm \AA$ around the Zr atoms is obtained, revealing that
the local structure differs notably from the average $\rm
Pm\bar3m$ cubic structure deduced from X-ray diffraction. In
particular, our results show that the distance between Zr atoms
and their first oxygen neighbors is independent of the Zr
substitution rate $x$ and equal to that measured in BaZrO$_3$,
while the X-ray cubic cell parameter increases linearly with $x$.
Furthermore, we show that the Zr atoms tend to segregate in
Zr-rich regions. We propose that the relaxor behavior in BTZ is
linked to random elastic fields generated by this particular
chemical arrangement, rather than to random electric fields as is
the case in most relaxors.

\end{abstract}

% EXAFS, ferroelectricity and antiferroelectricity, dielectric loss and relaxation, titanates.

\maketitle

\section{\label{Intro} Introduction}

The perovskite-type barium zirconate titanate,
BaTi$_{1-x}$Zr$_x$O$_3$ (BTZ), has attracted considerable
attention as a possible lead-free ferroelectric material to
replace the current industry standard lead-based ferroelectrics
\cite{SingleCryst,Poudre,DRAM1,DRAM2,BTZorder,BTZpiezoOrient}. The
particularity of BTZ is the continuous change of its properties
from a ferroelectric behavior at low Zr substitution rates to a
relaxor ferroelectric (relaxor) behavior at higher substitution
rates (0.25 $\le x \le$ 0.5) \cite{Annie}. The low substitution
regime has been mainly studied for its interest in lead-free
ferroelectric memories \cite{DRAM1,DRAM2}. However, due to the
recent report of outstanding electro-mechanical properties in
relaxor-based ferroelectric solid solutions
\cite{Park,LeadfreeGiant}, it is rather the relaxor regime of BTZ
which currently attracts a large research effort.

Relaxors are characterized by a broad and frequency-dependent
dielectric anomaly as a function of temperature, instead of a
sharp and frequency-independent divergence as in classical
ferroelectrics. In a relaxor such as PbMg$_{1/3}$Nb$_{2/3}$O$_{3}$
(PMN), the anomaly surprisingly occurs in a cubic average
structure which remains centrosymmetric (non polar) at all
temperatures. It is generally admitted that the peculiar
properties of relaxors are related to the presence of nano-scaled
polar regions \cite{Cross1,Cross2,Samara}, due to different cation
shifts in different parts of the structure.

Two necessary ingredients are often cited for a relaxor
ferroelectric: (i) the presence of Pb$^{2+}$ or Bi$^{3+}$ cations
(showing large displacements due to their lone-pair) and/or (ii) a
heterovalent cationic disorder (generating random local electric
fields \cite{Westphal} that break long-range polar correlations). They are both
present in the extensively studied model relaxor PMN (e. g. Refs.
\onlinecite{Burns,Bonneau,deMathan,PMN_RS,PMN_Diff_HP_Jens,PMN_Diff_neutrons,PMN_PDF}).
On the other hand, BTZ relaxors present a homovalent
Zr$^{4+}$/Ti$^{4+}$ substitution, which does not give rise to such
random electric fields. Therefore, another mechanism has to be
considered in order to account for the break of long-range
correlated displacements in BTZ. It has been proposed that random
elastic fields, induced by the difference in size of Zr$^{4+}$ and
Ti$^{4+}$ cations, play an important role \cite{REF}. Yet it is
not known how, and on which scale such random elastic fields build
up. The BaTi$_{1-x}$Zr$_x$O$_3$ solid solution thus opens an
interesting route towards the comprehension of the relaxor
behavior, and raises much interest
\cite{Annie,Farhi,RamanJens,Sciau,BTZtrans,BTZpression}.
Furthermore, it is intriguing that PbZr$_{1-x}$Ti$_x$O$_3$ (PZT)
shows no relaxation \cite{Noheda} while BTZ does.

The average crystal structure of the BTZ relaxors (0.25 $\le x
\le$ 0.5) is cubic (space group $\rm Pm\bar3m$) at any temperature
\cite{Sciau}. Local distortions away from the ideal cubic
structure expected from the observation of the relaxation behavior
are evidenced by the observation of a first-order Raman scattering
\cite{Farhi,RamanJens}, which is forbidden in a perfect primitive
cubic structure. However, the latter technique does not reveal the
nature of these distortions. The aim of the present study is to
determine the local structure and chemical order in BTZ relaxors.
For this purpose, we used EXAFS (Extended X-ray Absorption Fine
Structure) spectroscopy which is a structural and chemical local
probe. This technique has been proven to be a powerful tool to
analyze the local distortions in ferroelectric perovskites (e. g.
Refs. \onlinecite{KTN_XAFS_Temp,KNO_XAFS_P,PTO_XAFS_Temp,Cao}). On
the other hand, only a few EXAFS studies have been reported for
relaxors yet
\cite{PMN-XAFS,Relaxor-XAFS,XAFSmixed,XAFSPIN,KNBT-XAFS}.

In the following, we report an EXAFS study of three BTZ relaxors
($x$ = 0.25, 0.30 and 0.35) at the Zr K-edge. The local structure
parameters were determined up to distances of 4.5 {\AA} from the
Zr atoms, giving information on their local environment up to
their fourth neighbors. Hence, the Zr/Ti repartition could be
probed, as well as the distortions induced by the substitution of
Ti atoms by Zr ones in BaTiO$_3$. In order to have a proper
reference for the understanding of the local Zr environment, we
also studied BaZrO$_3$ under the same experimental conditions.

\section{\label{Exp} Experimental methods}

\subsection{\label{Samples} Samples and experiments}

BaTi$_{1-x}$Zr$_x$O$_3$ ($x$=0.25, 0.30, 0.35 and 1) powders were
synthesized  by solid state reaction, starting from the
appropriate amounts of BaCO$_3$, TiO$_2$, and ZrO$_2$ powders and
following the method described in Ref. \onlinecite{Annie}. They
were then characterized by X-ray diffraction at 300 K. All the
powders were found to be single phases within the accuracy of the
experiment. The cubic cell parameters are 4.054(1), 4.061(1),
4.072(1), and 4.192(1)~{\AA} for $x$ values of 0.25, 0.30, 0.35,
and 1 respectively.

EXAFS experiments at the Zr K-edge (17.998 keV) were carried out
on the FAME beamline BM30B at the European Synchrotron Radiation
Facility (ESRF). The X-ray absorption coefficient $\mu$ was
measured in the transmission mode as a function of the incident
photon energy $E$ in the range $\lbrack$ 17.8-19.5 keV $\rbrack$,
using a Si(220) single crystal monochromator. The samples
consisted of pellets made of BaTi$_{1-x}$Zr$_x$O$_3$ powders (with
a grain size of about 1 $\rm \mu$m) mixed with very low-absorbing
boron nitride. The effective thickness $e$ of each sample was
chosen in such a way that the product $\mu e$ increases
approximately by 1 through the Zr K-edge. EXAFS data were
collected at room temperature for all samples, as well as at 11 K
for BaZrO$_3$ and at 11K, 90 K and 150 K for
BaTi$_{0.65}$Zr$_{0.35}$O$_3$.

\subsection{\label{EXAFSintro} EXAFS data treatment}

\begin{figure}
\[\includegraphics[scale=0.4]{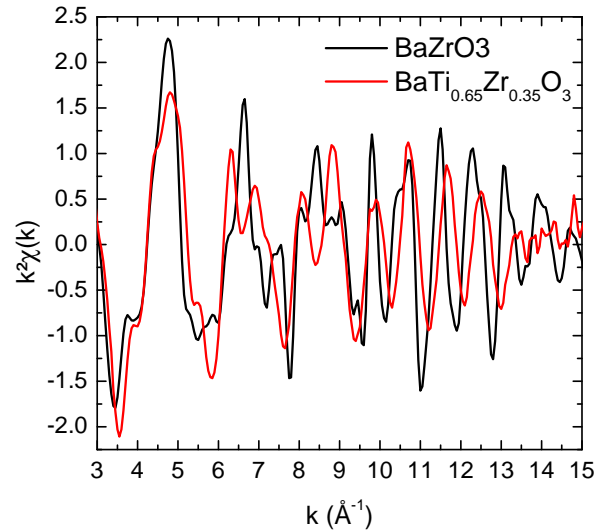}\]
\caption{ $k^2$-weighted normalized EXAFS signals, for BaZrO$_3$
and BaTi$_{0.65}$Zr$_{0.35}$O$_3$ at 300 K.} \label{chik300K}
\end{figure}

The scattering of the photoelectron by the neighbors of the
central absorbing atom introduces oscillations in the energy
dependence of the absorption coefficient $\mu (E)$ after the
energy edge. The normalized experimental EXAFS signal is given by
$\chi(k) = [\mu(k) - \mu_0(k)]/\Delta \mu_0 (k=0)$. $\mu_0$ is the
smooth, atomic-like, absorption background, $\Delta \mu_0 (k=0)$
is the absorption edge jump and $k$ is the photoelectron wave
number, given by $ k=\sqrt{2m_e(E-E_0)/{\hbar}^2}$ with $m_e$ the
electron mass and $E_0$ the edge energy. In the present work,
$\chi(k)$ was extracted from the measured absorption coefficient
$\mu (E)$, by using the AUTOBK program \cite{Autobk}. $E_0$ was
taken at the first maximum of the derivative of $\mu(E)$ at the Zr
K-edge. Two representative examples of normalized $k^2$-weighted
$\chi(k)$ spectra are shown in Fig.~\ref{chik300K}.

The theoretical EXAFS signal $\chi(k)$ is expressed as a sum of
contributions from different paths, each path $i$ corresponding to
a given scattering process of the photoelectron:
\begin{eqnarray}
\chi(k) & = & - {S_0}^2\sum_i \frac{N_iA_i(k)}{k{R_i}^2}
e^{-2k^2{\sigma_i}^2}e^{-2{R_i}/\lambda(k)} \nonumber \\ & &
\sin\left(2k{R_i}+2\delta_c(k)+\phi_i(k)\right), \label{equation1}
\end{eqnarray}
where $N_i$ is the degeneracy of path $i$, $R_i$ its half-length
and $A_i(k)$ its effective scattering amplitude. The Debye-Waller
(DW) factor ${\sigma_i}^2$ is the standard deviation of the $R_i$
distance distribution, assumed to be gaussian. The DW factor takes
into account both the thermal disorder and a possible small
structural disorder. $\delta_c(k)$ and $\phi_i(k)$ are phase
shifts associated with the electron propagating into and out the
potentials of the absorbing site and scattering sites
respectively. The other parameters are the photoelectron mean-free
path $\lambda(k)$ and an overall amplitude factor ${S_0}^2$, close
to 1, which accounts for many-electron effects in the excited
central atom. Note that Eq. \ref{equation1} includes both single
(back-)scattering (SS) and multiple scattering (MS) processes
\cite{Zabinsky}. For a SS path, $N_i$ is simply the number of
chemically identical atoms situated at a given distance $R_i$ from
the central atom.

The analysis of EXAFS signals at the Zr K-edge was performed in
$R$-space after a Fourier transform (FT) of the $k^2$-weighted
$\chi(k)$ in the $k$-range $\lbrack$ 3.2-14.6 \AA$^{-1} \rbrack$,
using a Hanning weight function. The calculated FTs were fitted to
the experimental ones using the FEFFIT program \cite{FEFFIT}. The
electronic parameters $A_i(k)$, $\phi_i(k)$, $\delta_c(k)$ and
$\lambda(k)$ were calculated for both SS and MS paths with the
FEFF8.00 code \cite{FEFF}. The structural parameters $R_i$,
${\sigma_i}^2$, as well as $N_i$ if unknown, were extracted from
the fit as will be described in details in sections \ref{shell1},
\ref{shell2BaZrO3}, and \ref{shell2BTZ}. As customary, an
additional parameter $\Delta E_0$ was introduced to account for
the small difference between the experimental edge energy $E_0$
and its calculated value using the FEFF code. The parameter
${S_0}^2$ was also refined, the experimental value usually
differing from the theoretical one.

\section{\label{Results} Results}

\subsection{\label{quali} General trends}

\begin{figure}
\[\includegraphics[scale=0.4]{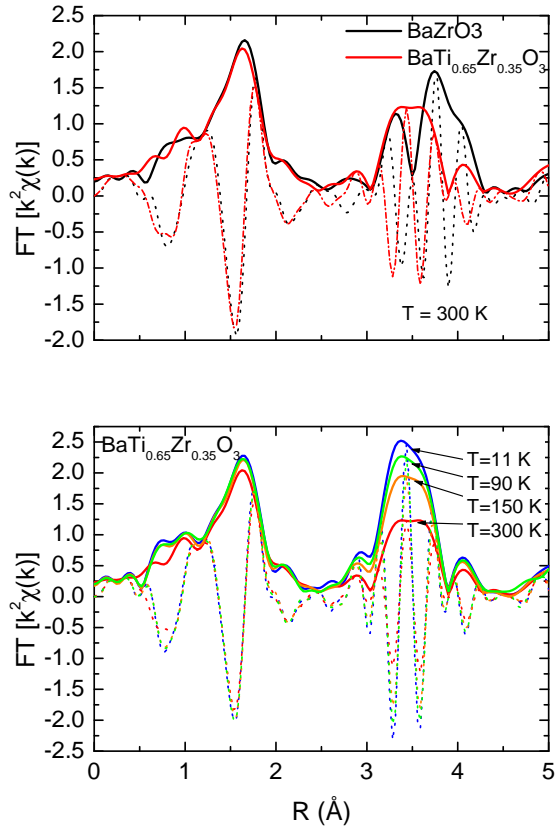}\]
\caption{Selected FTs of the $k^2$-weighted EXAFS signals:
BaZrO$_3$ and BaTi$_{0.65}$Zr$_{0.35}$O$_3$ at 300 K (top), and
thermal evolution in BaTi$_{0.65}$Zr$_{0.35}$O$_3$ (bottom). The
imaginary part and the modulus of the FTs are plotted as dashed
and solid lines respectively.} \label{CompTF}
\end{figure}

Examples of the Fourier transforms of the experimental
$k^2$-weighted $\chi(k)$ spectra are shown in Fig. \ref{CompTF}.
For all studied samples and at all temperatures, the FTs can be
readily separated into two contributions, characteristic of the
perovskite structure. The first neighbors of a Zr atom are six
oxygen atoms (denoted hereafter as O1), which form a regular
octahedron in the perfect $\rm Pm\bar3m$ perovskite structure .
The first contribution to the FT, at $R$ values lower than 2.5
\AA, only includes a SS process by one of the O1 atoms. The second
contribution, in the $R$-range $\lbrack$ 2.5-4.5 \AA $\rbrack$,
results both from SS processes where the photoelectron is
backscattered by either the second (Ba), third (Zr or Ti), or
fourth oxygen (O2) neighbors of the Zr atom, and from several MS
processes. Note that due to the $k$-dependence of the phase shifts
$\phi_i(k)$ and $\delta_c(k)$, the maxima of the FT modulus occur
at distances different from the real ones.

Surprisingly, the first peaks of the FTs are nearly identical at
300 K in all relaxor samples {\it and} in BaZrO$_3$ (see Figs.
\ref{CompTF} and \ref{fits1}). The latter suggests very close
first neighbors environment of the Zr atoms, whatever the Zr
substitution rate. On the other hand, the FTs of BaZrO$_3$ and BTZ
relaxors differ significantly for $R$ values larger than 2.5 \AA.

The shape of the FTs suggests an analysis in two steps. Fits of
the measured FTs are first performed in the $R$-range related to
the Zr first oxygen neighbors (Sec. \ref{shell1}). In a second
step, the fitted $R$-range is extended up to 4.5 $\rm\AA$, in
order to take into account the further neighbors contributions
(Sec. \ref{shell2BaZrO3} and \ref{shell2BTZ}). In BTZ relaxors,
one has to take into account both the Zr/Ti substitution and the
strong contribution of several MS paths in this $R$-range, which
makes the EXAFS analysis complicated. In order to get reference
parameters for the analysis of BTZ relaxors, the FT of BaZrO$_3$
is first fitted (Sec. \ref{shell2BaZrO3}). From previous EXAFS
studies \cite{Ravel,Bugaev,Petit}, the local structure of
BaZrO$_3$ can be considered as a perfect cubic perovskite
structure \cite{footnote1}, identical to the average one deduced
from X-ray diffraction (space group $\rm Pm\bar3m$)
\cite{Mathews,Ba-CaZrO3struct}.

\subsection{\label{shell1} Analysis of the first neighbor contribution}

\begin{figure}
\[\includegraphics[scale=0.4]{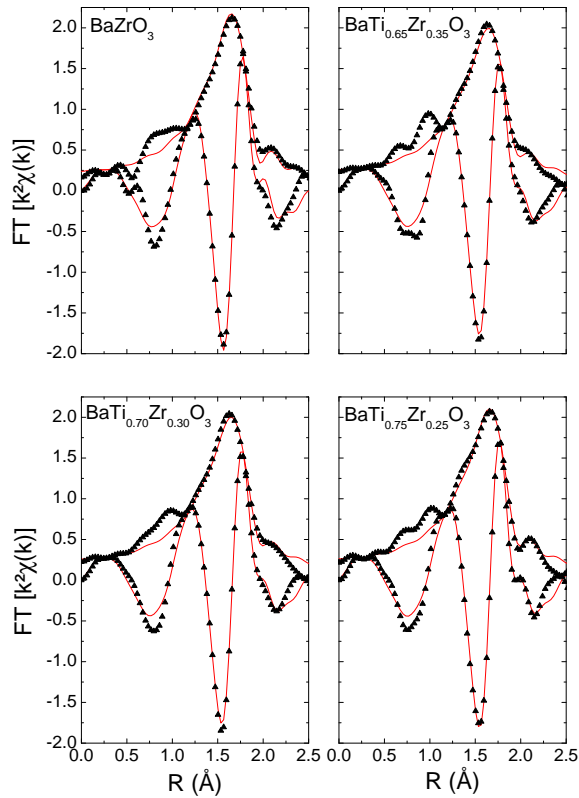}\]
\caption{Modulus and imaginary part of the FT of $k^2\chi(k)$ at
300 K. The dots represent measured data and the solid lines their
best fit, obtained with the parameters values given in Table
\ref{tableshell1}. Measured data for $R <$ 1 $\rm\AA$ are affected
by the background subtraction procedure and were excluded from the
analysis.} \label{fits1}
\end{figure}

We already noted that the backscattering processes between Zr
atoms and their first oxygen neighbors (O1) give rise to very
similar contributions to the FTs of the EXAFS signals below 2.5
\AA, for all samples and at all temperatures (Figs. \ref{CompTF}
and \ref{fits1}). From the known structure of BaZrO$_3$, it is
then reasonable to assume, as a starting point for the analysis of
BTZ relaxors, that the six O1 atoms are located at the same
distance from the absorbing Zr atom. Within this hypothesis, only
one SS path (denoted hereafter as path 1) contributes in
Eq.~\ref{equation1} for the considered $R$-range. The fitted
parameters are, for each sample, the Zr-O1 distance $R_1$, the
associated DW factor ${\sigma_1}^2$ and $\Delta E_0$. The number
of O1 neighbors, $N_1$, is fixed to 6. The analysis of the EXAFS
data for BaZrO$_3$ at 11 and 300 K allows to determine a ${S_0}^2$
value equal to 1 $\pm$ 0.07. This parameter was then fixed to 1
for all the further analysis. The EXAFS oscillations were refined
in the $R$-range $\lbrack$ 1.14-2.33 \AA $\rbrack$: the data for
$R$ lower than 1 $\rm\AA$ are affected by the background
subtraction procedure \cite{Autobk} and had to be excluded from
the fit. Fits of good quality could be obtained, with reliability
factors less than 1\%, for all samples at all temperatures (Fig.
\ref{fits1} and Table \ref{tableshell1}).

\begin{table}
\caption{Structural parameters deduced from the EXAFS analysis in
the $R$-range $\rm \lbrack$ 1.14-2.33 \AA $\rbrack$, for BaZrO$_3$
and BTZ relaxors. $R_1$ is the length of the bond between an
absorbing Zr atom and its first oxygen neighbor, and
${\sigma_1}^2$ the associated Debye-Waller factor. RF is the
reliability factor of the fit. The uncertainties are of the order
of $\pm$ 0.01 $\rm\AA$ and $\pm$ 0.0004 \AA$^2$ for $R_1$ and
${\sigma_1}^2$ respectively. The energy shift $\Delta E_0$ was
found equal to 1.4 $\pm$ 0.8 eV for all fits.} \label{tableshell1}
\begin{tabular}{ccccccc}
%\begin{tabular}{c|c|c|c|c|c}
\hline \hline sample & T & $R_1$ & ${\sigma_1}^2$ & RF
\\ & (K)& (\AA)& (\AA$^2$) & factor (\%) \\
\hline BaZrO$_3$ & \begin{tabular}{c} 300 \\ 11 \end{tabular} &
\begin{tabular}{c} 2.11 \\ 2.11 \end{tabular} &
\begin{tabular}{c} 0.0039 \\ 0.0029 \end{tabular} &
\begin{tabular}{c} 0.61 \\ 0.62 \end{tabular} \\
\hline BaTi$_{0.65}$Zr$_{0.35}$O$_3$ & \begin{tabular}{c} 300
\\ 150 \\ 90 \\ 11 \end{tabular} &
\begin{tabular}{c} 2.10 \\ 2.10 \\ 2.10 \\ 2.10 \end{tabular} &
\begin{tabular}{c} 0.0048 \\ 0.0038 \\ 0.0038 \\ 0.0035 \end{tabular} &
\begin{tabular}{c} 0.58 \\ 0.46 \\ 0.96 \\ 0.48 \end{tabular} \\
\hline BaTi$_{0.70}$Zr$_{0.30}$O$_3$ & 300 & 2.10 & 0.0048 & 0.55
\\ \hline BaTi$_{0.75}$Zr$_{0.25}$O$_3$ & 300 & 2.10 &
0.0043 & 0.47 \\ \hline\hline
\end{tabular}
\end{table}

In BaTi$_{0.65}$Zr$_{0.35}$O$_3$ and BaZrO$_3$ samples, no
significant temperature dependence of the Zr-O1 distance is
detected, within the experimental accuracy of the EXAFS technique.
More surprisingly, the Zr-O1 distance hardly varies with the Zr
substitution rate $x$, and keeps values very close to that found
in BaZrO$_3$.

\begin{figure}
\[\includegraphics[scale=0.4]{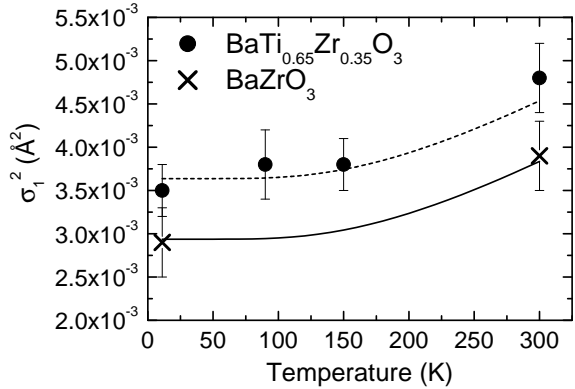}\]
\caption{Thermal evolution of the measured DW factor for the Zr-O1
bond (symbols). The solid line represents Eq. \ref{equation2} with
$\theta_E$ = 606 K. The dashed line corresponds to the same
function shifted by 0.0007 \AA$^2$.} \label{SigO1}
\end{figure}

The refined values of the ${\sigma_1}^2$ DW factors in BaZrO$_3$
at 11 and 300 K are in good agreement with those reported in Ref.
\onlinecite{Petit}. In BTZ relaxors, systematically higher
${\sigma_1}^2$ values are obtained. From the definition of the DW
factor, this increase depicts either a stronger static disorder,
or enhanced vibration amplitudes (associated with a decrease of
the Zr-O1 bond stiffness). As a matter of fact, the static and
dynamic contributions can be separated, by analyzing the thermal
evolution of the measured DW factor. In the absence of a static
disorder, the DW factor only accounts for thermal vibrations and
can be written, in the Einstein model, as:
\begin{eqnarray} {\sigma^2_{i~therm}}(T) =
{\frac {\hbar^2} {2 k_B M_R \theta_E}} \coth(\frac {\theta_E}
{2T}), \label{equation2}
\end{eqnarray}
where $k_B$ is the Boltzmann constant, and $M_R$ the reduced mass
of all atoms involved in the scattering path. The Einstein
temperature $\theta_E$, characterizing the bond strength,
increases with the bond stiffness. Given our hypothesis of a
perfect $\rm Pm\bar3m$ perovskite local structure in BaZrO$_3$,
the two ${\sigma_{1}}^2$ values measured at 11 and 300 K in this
sample were fitted to Eq. \ref{equation2}. An Einstein temperature
of 606 K $\pm$ 14 K was then obtained for the Zr-O1 bond. In
BaTi$_{0.65}$Zr$_{0.35}$O$_3$, the thermal evolution of
${\sigma_1}^2$ is consistent with the same $\theta_E$ value,
provided that a constant ${\Delta\sigma_1}^2$ = + 0.0007 {\AA}$^2$
is added to Eq. \ref{equation2} (Fig. \ref{SigO1}). This result
gives evidence for the existence of a static distribution of the
Zr-O1 distances in BaTi$_{0.65}$Zr$_{0.35}$O$_3$, which is
temperature-independent within the experimental accuracy. The full
width at half maximum of a gaussian distribution of distances
would be equal to $2\sqrt{2\ln2{\Delta\sigma_1}^2}$ = 0.06 \AA.
This small value justifies {\it a posteriori} the use of only one
SS path in Eq. \ref{equation1}, the DW factor ${\sigma_1}^2$
taking into account the small disorder that cannot be resolved in
$R$-space.

The mean square deviation of the $R_i$ distance linked to a static
disorder, ${\Delta\sigma_i}^2$, is intended to describe a gaussian
static distribution of distances in Eq. \ref{equation1}, with
${\sigma_i}^2 = {\sigma^2_{i~therm}}(T) + {\Delta\sigma_i}^2$.
However, a measured non-zero value of ${\Delta\sigma_i}^2$ can
describe other distance distributions (discrete or continuous),
provided their width remains small. In the present case, the
distribution of the Zr-O1 distances could be due to a distortion
of the ZrO$_6$ units, coming from the Zr/Ti chemical disorder,
and/or to a displacement of Zr atoms in their octahedral cages. To
evaluate the magnitude of such a displacement, we calculated the
FT of the EXAFS signals for Zr atoms in a perfect octahedron,
weakly displaced either along the [100], [110], or [111] cubic
axis. We then performed fits on the calculated signals in the same
conditions as described above for the measured signals. From these
simulations, we conclude that the refined DW factors presented in
Table \ref{tableshell1} for BaTi$_{0.65}$Zr$_{0.35}$O$_3$ could
correspond to a Zr displacement of 0.07 \AA, in any of the three
directions cited above.

A static distribution of Zr-O1 distances is also expected in the
two other relaxor samples, BaTi$_{0.70}$Zr$_{0.30}$O$_3$ and
BaTi$_{0.75}$Zr$_{0.25}$O$_3$, from the ${\sigma_1}^2$ values
measured at 300~K and by analogy with the case of
BaTi$_{0.65}$Zr$_{0.35}$O$_3$.

In conclusion, the length and strength of the bond between a Zr
atom and its first oxygen neighbor are found to be independent of
the Zr substitution rate in BaTi$_{1-x}$Zr$_{x}$O$_3$ samples.
Only a small temperature-independent distribution of the Zr-O1
distances is detected in BaTi$_{0.65}$Zr$_{0.35}$O$_3$, in
opposition to the single distance in the regular octahedron
present in BaZrO$_3$. These fluctuations of the Zr-O1 distance can
be due to Zr displacements in their octahedra, but can also result
from distortions of the ZrO$_6$ octahedra induced by the Zr/Ti
chemical disorder.

\subsection{\label{shell2BaZrO3} Analysis of the further neighbor contribution: BaZrO$_3$}

For $R$ values up to 4.5 \AA, the FT of the EXAFS signal involves
scattering processes beyond the first oxygen neighbors. The
amplitudes and phases of all the possible paths have been
calculated by using the FEFF8.00 code, for an atomic cluster of
145 atoms representative of the BaZrO$_3$ $\rm Pm\bar3m$
perovskite structure. The scattering processes with relative
weight lower than 2.5 \% have been neglected. The single
backscattering processes on the second (Ba), third (Zr), and
fourth (O2) neighbors of the Zr central atom must be considered,
as well as several collinear MS paths and only two non-linear MS
paths. One of the two latter paths involves Ba atoms. The other
one is a triple scattering path within the ZrO$_6$ octahedron,
which corresponds to a $R$-range between the two main peaks of the
FTs. We verified that it does not affect the fit, and did not
consider it further. The retained paths for the EXAFS analysis are
listed in Table \ref{tablepaths}. Note that the half-lengths of
all the collinear MS paths are equal to the Zr-Zr distance. As a
consequence, these MS paths contribute to the FT in the same
$R$-range as the single backscattering path by a Zr atom, which
makes the analysis complex.

\begin{table}
\caption{SS and MS paths used to analyze BaZrO$_3$ EXAFS data.
$N_i$ is the degeneracy of path $i$, and $R_i$ its half-length,
expressed as a function of $R_1$ assuming an ideal perovskite
structure. $\rm Zr_c$ is the central absorbing Zr atom. The O1 and
O1' atoms are located on opposite sides of the $\rm Zr_c$ atom,
between $\rm Zr_c$ and one of its six Zr third neighbors.}
\label{tablepaths}
\begin{tabular}{c|c|c|c|c}
\hline \hline  index & scattering process & $N_i$ & $R_i$ &
${\sigma_i}^2$ \\ \hline 1 & $\rm  Zr_c \rightarrow  O1
\rightarrow Zr_c$ & 6 & $R_1$ & ${\sigma_1}^2$
\\ \hline 2 &  $\rm  Zr_c \rightarrow  Ba  \rightarrow
Zr_c$ & 8 & $\sqrt{3}R_1$ & ${\sigma_2}^2$
\\ \hline 3 & $\rm  Zr_c \rightarrow  Zr  \rightarrow
Zr_c$ & 6 & $2R_1$ & ${\sigma_3}^2$
\\ \hline 4 & $\rm  Zr_c \rightarrow  O1'  \rightarrow  O1 \rightarrow
Zr_c$ & 6 & $2R_1$ & ${\sigma_4}^2$
\\ \hline 5 & $\rm  Zr_c \rightarrow  Zr \rightarrow  O1  \rightarrow
Zr_c$ & 12 & $2R_1$ & ${\sigma_3}^2$
\\ \hline 6 &  $\rm  Zr_c \rightarrow  O1'  \rightarrow Zr_c  \rightarrow  O1 \rightarrow
Zr_c$ & 6 &  $2R_1$ & ${\sigma_4}^2$
\\ \hline 7 &  $\rm  Zr_c \rightarrow  O1  \rightarrow Zr_c  \rightarrow  O1 \rightarrow
Zr_c$ & 6 & $2R_1$ & 2${\sigma_1}^2$
\\ \hline 8 &  $\rm  Zr_c \rightarrow  O1  \rightarrow Zr  \rightarrow  O1 \rightarrow
Zr_c$ & 6 & $2R_1$ & ${\sigma_3}^2$
\\ \hline 9 &  $\rm  Zr_c \rightarrow  O1  \rightarrow Ba \rightarrow
Zr_c$ & 48  & $\frac{1+\sqrt{2}+\sqrt{3}}{2}R_1$ & ${\sigma_9}^2$
\\ \hline 10 &  $\rm  Zr_c \rightarrow  O2  \rightarrow
Zr_c$ & 24  & $\sqrt{5}R_1$ & ${\sigma_{10}}^2$
\\ \hline \hline
\end{tabular}
\end{table}

Assuming a perfect $\rm Pm\bar3m$ cubic structure, all $N_i$ are
known and the $R_i$ parameters for the ten paths can be expressed
as a function of $R_1$, the distance between the central absorbing
Zr atom and O1 (see Table \ref{tablepaths}). The number of
${\sigma_i}^2$ parameters needed could be decreased using the
expression of the DW factor \cite{Zabinsky}:
\begin{eqnarray}
{\sigma_i}^2 = \frac 1 4 \langle [ \sum_j
(\overrightarrow{u}_{j+1}-\overrightarrow{u}_j).\overrightarrow{R}_{j
j+1} ]^2 \rangle_t, \label{equation3}
\end{eqnarray}
where $t$ is the time, $j$ represents an atomic site of the
scattering path $i$, $\overrightarrow{u}_j$ is the displacement
vector of the $j$ atom, and $\overrightarrow{R}_{j j+1}$ is the
directing unit vector between $j$ and $j+1$ atoms at equilibrium.
The resulting relations between the ${\sigma_i}^2$ parameters in
BaZrO$_3$ are given in Table \ref{tablepaths}. For the fit of the
measured FT, in the $R$-range $\lbrack$ 1.14-4.52~\AA $\rbrack$,
seven parameters were then refined: $\Delta E_0$, $R_1$,
${\sigma_2}^2$, ${\sigma_3}^2$, ${\sigma_{4}}^2$,
${\sigma_{9}}^2$, and ${\sigma_{10}}^2$. Although they have
already been determined in Sec. \ref{shell1}, $R_1$ and the
strongly correlated parameter $\Delta E_0$ were refined
\cite{footnote2}, all distances up to 4.5 $\rm \AA$ being
functions of $R_1$. On the other hand, the ${S_0}^2$ and
${\sigma_1}^2$ parameters were fixed to the previously refined
values. The obtained parameters at 11 K and 300 K are presented in
Table \ref{fitBZ}. During the refinement process, the
${\sigma_{4}}^2$ and ${\sigma_{9}}^2$ values were found to have no
influence on the RF factor. These two DW factors are related to
paths 4, 6, and 9, which give rise to very large, and relatively
weak contributions in $R$-space. Consequently, ${\sigma_{4}}^2$
and ${\sigma_{9}}^2$ values do not affect significantly the
amplitude and shape of these contributions and cannot be precisely
determined, so that they are not given in Table \ref{fitBZ}.

\begin{figure}
\[\includegraphics[scale=0.4]{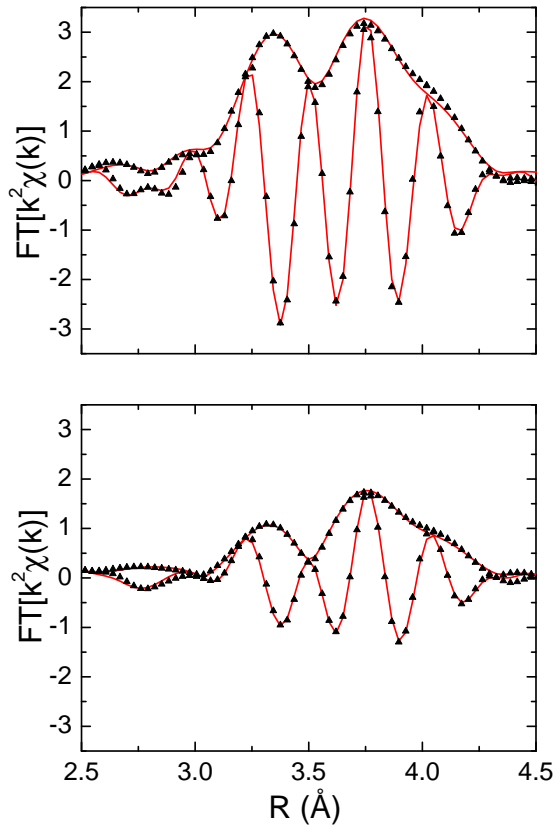}\]
\caption{Modulus and imaginary part of the FT of $k^2\chi(k)$ for
BaZrO$_3$, at 11 K (top) and  300 K (bottom). The dots represent
measured data and the solid lines their best fit. The fit
parameters are given in Table \ref{fitBZ}.} \label{fit2BZ}
\end{figure}

\begin{table}
\caption{Structural parameters deduced from the EXAFS analysis in
the $R$-range $\rm \lbrack$ 1.14-4.52 \AA $\rbrack$, for
BaZrO$_3$. The parameters have already been defined in Tables
\ref{tableshell1} and \ref{tablepaths}. The uncertainties are of
the order of $\pm$ 0.01 $\rm \AA$ for $R_1$, $\pm$ 0.0003 $\rm
\AA^2$ for ${\sigma_2}^2$ and ${\sigma_3}^2$, and $\pm$ 0.003 $\rm
\AA^2$ for ${\sigma_{10}}^2$. $\Delta E_0$ was found equal to 0.0
$\pm$ 0.5 eV for both fits.}\label{fitBZ}
\begin{tabular}{cccccc}
\hline \hline & $R_1$ & ${\sigma_2}^2$ & ${\sigma_3}^2$ &
${\sigma_{10}}^2$ & RF
\\
 & ($\rm \AA$) & ($\rm \AA^2$)& ($\rm \AA^2$) & ($\rm \AA^2$) & factor (\%)
\\
\hline 300 K & 2.10 & 0.0072 & 0.0048 & 0.013 & 1.04
\\
\hline 11 K & 2.10 & 0.0026 & 0.0025 & 0.006 & 1.01
\\
\hline\hline
\end{tabular}
\end{table}

The calculated FTs are in very good agreement with the measured
ones (Fig. \ref{fit2BZ}). The DW factors for the Zr-Ba
(${\sigma_2}^2$), Zr-Zr (${\sigma_{3}}^2$) and Zr-O$_2$
(${\sigma_{10}}^2$) bonds are consistent with those reported in
Ref. \onlinecite{Petit}. Their thermal evolution was fitted to Eq.
\ref{equation2}, resulting in Einstein temperatures equal to 194
$\pm$ 8 K, 263 $\pm$ 10 K, and 296 $\pm$ 50 K for the Zr-Ba,
Zr-Zr, and Zr-O2 bonds respectively.

\begin{figure}
\[\includegraphics[scale=0.4]{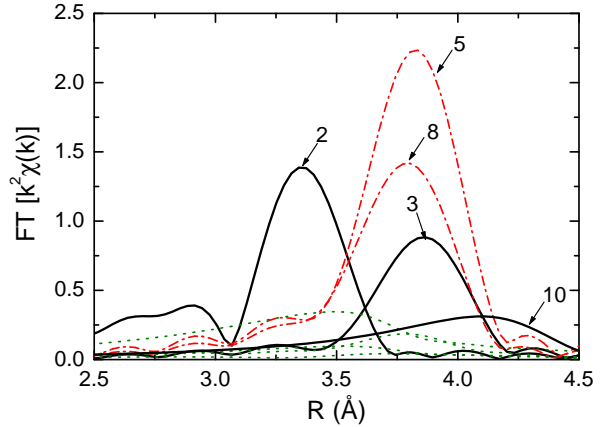}\]
\caption{Modulus of the FT of the individual contributions to
$k^2\chi(k)$, in BaZrO$_3$ at 300 K. SS paths, MS paths within
Zr$\rm_c$ octahedra, and MS paths outside Zr$\rm_c$ octahedra are
represented as solid, dash-dotted, and dotted lines respectively.
Numbers correspond to the path indexes of Table \ref{fitBZ}.}
\label{CompIndPaths}
\end{figure}

The analysis of the EXAFS oscillations in BaZrO$_3$ gives
information on the relative contributions of the different paths
to the FTs in the $R$-range $\lbrack$~2.5-4.5 \AA $\rbrack$, which
will help to analyze BTZ samples (see Fig. \ref{CompIndPaths}).
Only the backscattering path on Ba atoms (path 2 in Table
\ref{tablepaths}) contributes in the $R$-range [3.0-3.6 \AA],
which allows a precise determination of the Zr-Ba distance and the
associated DW factor. The contributions in the $R$-range [3.6-4.2
\AA] mainly arise from the MS linear paths that involve the third
neighbor Zr atoms (paths 5 and 8 in Table \ref{tablepaths}). The
backscattering path on Zr atoms (path 3) also contributes in the
same $R$-range, but it is much less important. The remaining MS
paths (paths 4, 6, 7, and 9 in Table \ref{tablepaths}) correspond
to weak contributions to the FTs, but it is necessary to take them
into account in order to obtain good quality fits. The DW factors
of paths 4, 6, and 9 were shown to have no influence on the fit.
Finally, the backscattering path on O2 atoms (path 10) contributes
over a very large $R$-range [3-4.6 \AA]. It is then compulsory to
keep this path in the analysis, as it cannot be separated from the
other contributions to the FTs.

\subsection{\label{shell2BTZ} Analysis of the further neighbor
contribution: BaTi$_{1-x}$Zr$_{x}$O$_3$ relaxors}

In the following analysis, we shall introduce two different models
of the local structure in BTZ relaxors. In a first step, we show
that a model with aligned Zr, O1, and (Zr/Ti) atoms in the
perovskite structure cannot account for the measured EXAFS
signals. In a second step, we introduce a buckling angle for the
Zr-O1-(Zr/Ti) bonds.

\subsubsection{\label{model1} Basic model for the BTZ analysis}

The BaZrO$_3$ model (see Sec. \ref{shell2BaZrO3}) is used as a
starting point to analyze BTZ samples. To take into account the Ti
third neighbors, we also calculated the amplitudes and phases of
the scattering paths within a hypothetical cluster of BaTiO$_3$ in
the $\rm Pm\bar3m$ cubic perovskite structure, with a single Zr
impurity as the absorbing atom. For the paths that do not involve
Ti atoms, the calculated amplitude and phase remain strictly
identical to those calculated for the BaZrO$_3$ structure. Thus,
we only consider three extra paths 3' ($\rm Zr_c \rightarrow  Ti
\rightarrow Zr_c$), 5' ($\rm Zr_c \rightarrow  Ti \rightarrow  O1
\rightarrow Zr_c$), and 8' ($\rm Zr_c \rightarrow O1 \rightarrow
Ti \rightarrow  O1 \rightarrow Zr_c$), in addition to those
presented in Table \ref{tablepaths}. The $k$-dependence of these
path amplitudes differs significantly from that of the
corresponding paths 3, 5, and 8 of the BaZrO$_3$ model. Thus, the
EXAFS technique allows the refinement of structural parameters for
Zr and Ti atoms with limited correlations effects, despite similar
Zr-Zr and Zr-Ti distances. A new parameter $N_{Zr}$ has to be
introduced, which is the average number of Zr third neighbors of
the central atom Zr$\rm _c$. The degeneracy of paths 3, 5, and 8
is then multiplied by $N_{Zr}$/6, and that of paths 3', 5', and 8'
by (6-$N_{Zr}$)/6. A unique ${\sigma_{3'}}^2$ parameter is
attached to the paths 3', 5' and 8', according to Eq.
\ref{equation3}.

Zr atoms being larger than Ti ones, they likely cause local
distortions in the BaTiO$_3$ matrix. This difference in size is
taken into account by introducing independent Zr-O1, Zr-Ba, Zr-Ti,
and Zr-O2 distances. The O1 atoms lying between Zr$\rm_c$ and
(Zr/Ti) atoms in this first model, the Zr-Zr and Ti-O1 distances
are defined using the expressions $d_{Zr-Zr} = 2 d_{Zr-O1}$ and
$d_{Ti-O1} = d_{Zr-Ti} - d_{Zr-O1}$. The DW factors ${\sigma_4}^2$
and ${\sigma_9}^2$ were fixed to the same values as in BaZrO$_3$,
since they are not expected to affect the fit (see Sec.
\ref{shell2BaZrO3}). Furthermore, the relations between the DW
factors presented in table \ref{tablepaths} remain correct, so
that only ${\sigma_2}^2$, ${\sigma_3}^2$, ${\sigma_{3'}}^2$, and
${\sigma_{10}}^2$ DW factors were refined. The Zr-O1 distance
($R_1$) and ${\sigma_1}^2$, as well as $\Delta E_0$ and ${S_0}^2$,
were fixed to the values refined in the $R$-range [1.14-2.33 \AA]
(see Table \ref{tableshell1}).

\begin{figure}
\[\includegraphics[scale=0.4]{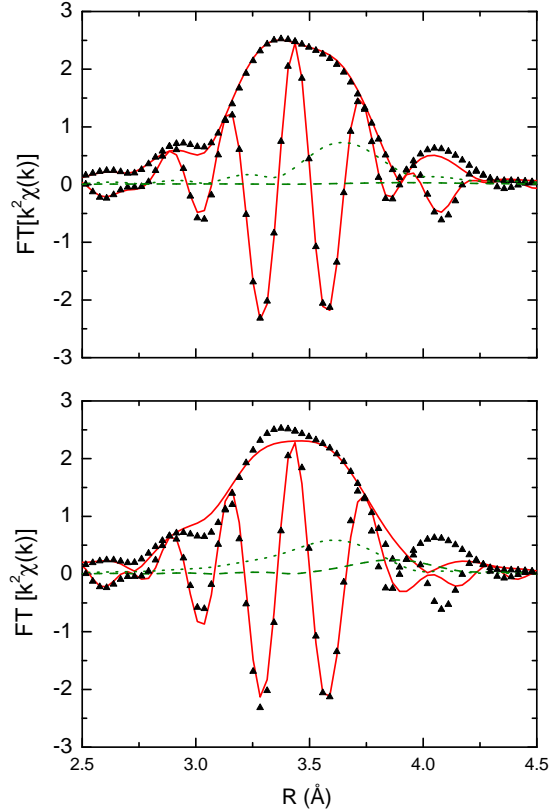}\]
\caption{Modulus and imaginary part of the FTs of the observed
(dots) and refined (solid line) $k^2\chi(k)$ for
BaTi$_{0.65}$Zr$_{0.35}$O$_3$ at 11 K. The modulus of the FT for
paths 3 (dashed line) and 3' (dotted line), which describe the
backscattering on Zr and Ti atoms respectively, are also
represented. The fits were obtained in the hypothesis of aligned
Zr$\rm _c$, O1 and (Zr/Ti) atoms. Top: the path 3 is completely
damped by a non-physical, huge ${\sigma_3}^2$ DW factor. Bottom:
${\sigma_3}^2$ is fixed to its value in BaZrO$_3$, i.e. 0.0020
\AA$^2$. The calculated imaginary part is then shifted towards
high $R$-values in the $R$-range [3.7-3.9 \AA], where the path 3
now contributes. The latter clearly indicates an overestimated
$d_{Zr-Zr}$ distance.} \label{FitNeg1}
\end{figure}

Within this model, the fitting procedure over the $R$-range
[1.14-4.52 \AA] yields non-physical values of the ${\sigma_{3}}^2$
and ${\sigma_{3'}}^2$ parameters (0.033 and -0.0008 respectively,
at 11 K). In fact, the calculated contribution of paths linked to
Zr atoms (paths 3, 5, and 8 in Table \ref{tablepaths}) are
completely damped (see top of Fig. \ref{FitNeg1}). Only paths
involving Ti atoms then contribute to the calculated signal, in
the $R$-range [3.4-3.9 $\rm\AA$]. Fixing ${\sigma_{3}}^2$ to its
value in BaZrO$_3$ made impossible a good agreement between
calculated and experimental signals. In particular, the shift of
the FT's imaginary part in the $R$-range [3.6-3.9 \AA] indicates
an overestimated Zr-Zr distance (see bottom of Fig.
\ref{FitNeg1}). Hence, we deduce that $d_{Zr-Zr}$ must be smaller
than $2d_{Zr-O1}$.

\subsubsection{\label{model2} Determination of the buckled local structure in BTZ relaxors}

\begin{figure}
\[\includegraphics[scale=0.4]{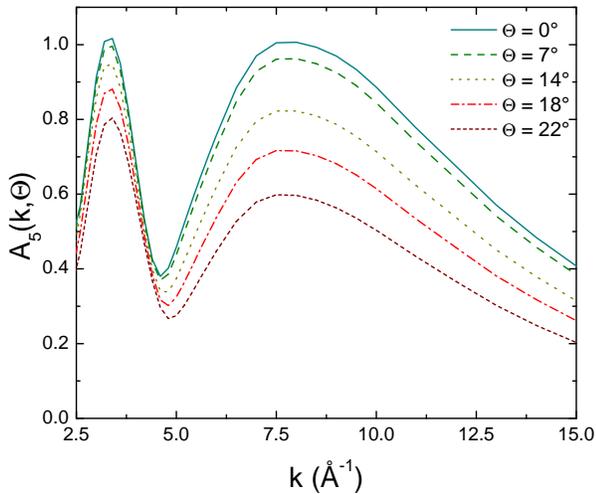}\]
\caption{Evolution of the focusing path amplitudes with increasing
values of the buckling angle $\Theta$: example of path 5.}
\label{Ampl6}
\end{figure}

The hypothesis of aligned Zr$\rm_c$, O1 and Zr atoms being not
valid in BTZ samples, a new model has to be built, where the O1
atoms are no longer aligned with Zr ones. In order to get all
different possible configurations, one can introduce two mean
buckling angles, $\Theta_{Zr}$ and $\Theta_{Ti}$, defined as
180$\rm^o$ - $\rm\widehat{Zr O1 Zr}$ and 180$\rm^o$ -
$\rm\widehat{Zr O1 Ti}$ respectively. All the possible paths and
their parameters were calculated in the preceding BaZrO$_3$ and
BaTiO$_3$ clusters, moving the O1 atoms out of their collinear
sites with varying values of the buckling angle $\Theta$. The only
effect of the O1 atom displacements is the damping of the
amplitudes of the MS paths in which these atoms play a focusing
role, i.e. paths 5, 5', 8, and 8' (Table \ref{tablepaths}). As an
example, we show the evolution of the amplitude of path 5 as a
function of $\Theta$ on Fig. \ref{Ampl6}. This evolution can be
expressed by expending the scattering amplitude of the focusing
paths about $\Theta$ = 0:
\begin{eqnarray}
A(k,\Theta) \approx {A(k,0)[1-b(k)\Theta^2]}^n, \label{equation4}
\end{eqnarray}
where n is the number of scattering processes by the off-centered
focusing atom (O1) \cite{Frenkel}. For each of the paths 5, 5', 8,
and 8', the coefficients $b(k)$ were determined for various
buckling angles and averaged. Eq. \ref{equation4} is found to be
correct for buckling angles values below 20$\rm^o$.

\begin{figure}
\[\includegraphics[scale=0.4]{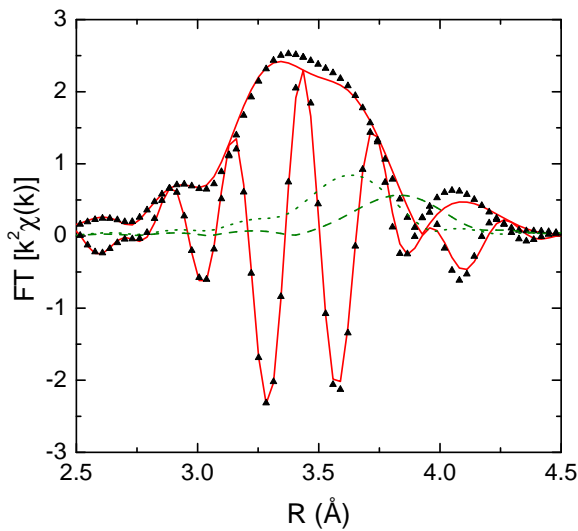}\]
\caption{Modulus and imaginary part of the FTs of the observed
(dots) and refined (solid line) $k^2\chi(k)$ for
BaTi$_{0.65}$Zr$_{0.35}$O$_3$ at 11 K. The modulus of the FT for
paths 3 (dashed line) and 3' (dotted line), which describe the
backscattering on Zr and Ti atoms respectively, are also
represented. We show the best fit obtained in the hypothesis of a
single buckling angle for Zr$\rm _c$-O1-Zr and Zr$\rm _c$-O1-Ti
bonds, for reasonable values of ${\sigma_{3}}^2$ and
${\sigma_{3'}}^2$ parameters. The missing amplitude in the
$R$-range [3.3-3.7 \AA], characteristic of the path 3', indicates
an overestimated buckling angle for the Ti bonds.} \label{FitNeg2}
\end{figure}

\begin{figure}
\[\includegraphics[scale=0.4]{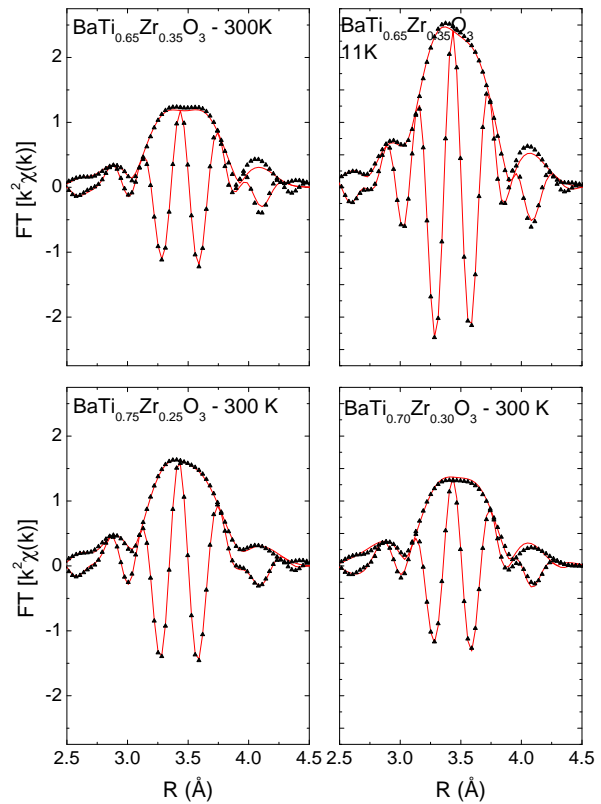}\]
\caption{Modulus and imaginary part of the FT of $k^2\chi(k)$ at
either 11 K or 300 K, for BTZ samples. The dots represent measured
data and the solid lines their best fit, obtained with the
parameters values given in Table \ref{fitBTZ}.} \label{fit2BTZ}
\end{figure}

The expressions for Zr-Zr and Ti-O1 distances are now given by:
$d_{Zr-Zr} = 2 d_{Zr-O1} \cos(\Theta_{Zr}/2)$, and $d_{Ti-O1} =
\sqrt{d_{Zr-Ti}^2 - d_{Zr-O1}^2 \sin^2\Theta_{Ti}} - d_{Zr-O1}
\cos\Theta_{Ti}$, from which the lengths of paths 1 to 8 can be
easily derived. The path length $R_9$ splits into several
different values, due to the misalignment of the O1 oxygen atoms.
This effect can be modeled by an increase of the corresponding DW
factor ${\sigma_9}^2$, but we already showed that this parameter
does not affect the fit quality. Thus, we determined the mean
distance $R_9$ with the expression given in Table
\ref{tablepaths}, using the the ${\sigma_9}^2$ value of BaZrO$_3$.
The length of path 10 splits as well, depending on the size and
shape of the octahedra next to the central one Zr$\rm _c$O$_6$.
This information being out of the scope of this study, we assigned
to this path an average value of the Zr-O2 distances, $R_{10}$. In
that case, ${\sigma_{10}}^2$ is expected to take into account both
thermal and static disorders. In the absence of an adequate model
for the O2 repartition, the values obtained for $R_{10}$ and
${\sigma_{10}}^2$ will not be interpreted. Concerning the DW
factors, the linearity of paths 5, 5', 8 and 8' is broken by the
off-centering of O1 atoms with respect to Zr-(Zr/Ti) bonds, so
that the relations ${\sigma_5}^2$ = ${\sigma_8}^2$ =
${\sigma_{3}}^2$ and ${\sigma_5'}^2$ = ${\sigma_8'}^2$ =
${\sigma_{3'}}^2$ are no longer exact. However, we postulated
their validity for the buckled paths, in order to limit the number
of free parameters. The buckling angles $\Theta_{Zr}$ and
$\Theta_{Ti}$ were determined separately, as preliminary fits
using the same buckling angle for both Zr$\rm _c$-O1-Zr and Zr$\rm
_c$-O1-Ti bonds did not yield a good agreement between
experimental and calculated signals (see Fig. \ref{FitNeg2}). Ten
parameters were then refined: $R_{2}$, $R_{3'}$, $R_{10}$,
$N_{Zr}$, $\Theta_{Zr}$, $\Theta_{Ti}$, ${\sigma_{2}}^2$,
${\sigma_{3}}^2$, ${\sigma_{3'}}^2$, and ${\sigma_{10}}^2$. The
${\sigma_{3}}^2$ and ${\sigma_{3'}}^2$ parameters were found to
reach very low (non-physical) values when not fixed. Like
$N_{Zr}$, $\Theta_{Ti}$, and $\Theta_{Zr}$, these two DW factors
determine the signal amplitude in the $R$-range [~3.4-3.9 \AA ].
Strong correlation effects exist between these five parameters and
various sets of their values give rise to similar, very good fits.
To go further in the analysis, we assume that ${\sigma_{3}}^2$ and
${\sigma_{3'}}^2$ in BTZ relaxors are equal to ${\sigma_{3}}^2$ in
BaZrO$_3$, which is determined for all temperatures using Eq.
\ref{equation2} with the Einstein temperature of 263 K reported in
Section \ref{shell2BaZrO3}. Values of refined parameters in this
hypothesis are presented in Table \ref{fitBTZ} and fits are shown
on figure \ref{fit2BTZ} for the three BTZ relaxors. The fits are
not very sensitive to the variations of $\Theta_{Ti}$ in the range
[0-10$\rm^o$]. $\Theta_{Ti}$ values larger than 10$\rm^o$ are
excluded since they do not yield satisfactory agreements. Note
that the Zr-Ba distance and the associated DW factor
${\sigma_{2}}^2$ are accurately defined, the contribution of path
2 to the FTs being well separated from the others.

One needs to estimate the consequences of fixing ${\sigma_{3}}^2$
and ${\sigma_{3'}}^2$ for the analysis of BTZ relaxors. For this
purpose, we performed several fits for various, not necessarily
equal, values of ${\sigma_{3}}^2$ and ${\sigma_{3'}}^2$, in a
range of $\pm$ 0.0010 $\rm \AA^2$ around the ${\sigma_{3}}^2$
value in BaZrO$_3$. The $\Theta_{Zr}$, $N_{Zr}$, and $R_{3'}$
values were found to vary by $\pm$ 1.5$\rm^o$, $\pm$ 0.2, and
$\pm$ 0.01 $\rm \AA$ respectively. Furthermore, the refined values
of $\Theta_{Zr}$ being at the limit of the validity range of Eq.
\ref{equation4}, one expects an error of $\pm$ 1$\rm^o$ on its
determination. The uncertainties given in Table \ref{fitBTZ} are
those determined by the FEFFIT code, augmented by the preceding
amounts. Note that in the case of BaTi$_{0.65}$Zr$_{0.35}$O$_3$,
the refined value of $N_{Zr}$ is found to be independent of the
temperature, as it should be. This confirms {\it a posteriori}
that the hypothesis of an identical thermal evolution of
${\sigma_{3}}^2$ and ${\sigma_{3'}}^2$ in BTZ samples and
${\sigma_{3}}^2$ in BaZrO$_3$ is reasonable.

\begin{table*}
\caption{Structural parameters deduced from the EXAFS analysis in
the $R$-range $\rm \lbrack$ 1.14-4.52 \AA $\rbrack$, for BTZ
relaxors. The path half-lengths are denoted $R_i$, and the
associated DW factors ${\sigma_i}^2$. $\Theta_{Zr}$ is the
buckling angle of the Zr-O1-Zr bonds, defined as
180$\rm^o$~-~$\rm\widehat{Zr O1 Zr}$, and $N_{Zr}$ the mean number
of Zr third neighbors of the Zr$\rm _c$ atom. $R_3$ is deduced
from $R_1$ (Table \ref{tableshell1}) and $\Theta_{Zr}$. The
uncertainties are of the order of $\pm$ 0.006 \AA, $\pm$ 0.0004
\AA$^2$, $\pm$ 5$\rm^o$, $\pm$ 0.04 \AA, $\pm$ 0.02 \AA, $\pm$
0.04 \AA, $\pm$ 0.005 \AA$^2$, and $\pm$ 1.0 for $R_2$,
${\sigma_2}^2$, $\Theta_{Zr}$, $R_3$, $R_{3'}$, $R_{10}$,
${\sigma_{10}}^2$ and $N_{Zr}$ respectively.} \label{fitBTZ}
\begin{tabular}{ccccccccccc}

\hline \hline

sample & T & $R_2$ & ${\sigma_2}^2$ & $\Theta_{Zr}$ & $R_3$ &
$R_{3'}$ & $R_{10}$ & ${\sigma_{10}}^2$ & $N_{Zr}$ & RF \\

 & (K) & ($\rm \AA$) & ($\rm \AA^2$) & ($\rm^o$) & ($\rm \AA$) & $(\rm
\AA$) & ($\rm \AA$) & ($\rm \AA^2$) &  & factor (\%) \\

\hline

BaTi$_{0.65}$Zr$_{0.35}$O$_3$ &
\begin{tabular}{c} 300 \\ 150 \\ 90 \\ 11 \end{tabular} &
\begin{tabular}{c} 3.557 \\ 3.554 \\ 3.556 \\ 3.554 \end{tabular} &
\begin{tabular}{c} 0.0070 \\ 0.0048 \\ 0.0041 \\ 0.0037 \end{tabular} &
\begin{tabular}{c} 18 \\ 17 \\ 19 \\ 18 \end{tabular} &
\begin{tabular}{c} 4.15 \\ 4.15 \\ 4.15 \\ 4.14 \end{tabular} &
\begin{tabular}{c} 4.07 \\ 4.06 \\ 4.07 \\ 4.06 \end{tabular} &
\begin{tabular}{c} 4.61 \\ 4.60 \\ 4.60 \\ 4.60 \end{tabular} &
\begin{tabular}{c} 0.018 \\ 0.014 \\ 0.012 \\ 0.012 \end{tabular} &
\begin{tabular}{c} 3.4 \\ 3.4 \\ 3.4 \\ 3.3 \end{tabular} &
\begin{tabular}{c} 1.09 \\ 0.81 \\ 0.80 \\ 0.51 \end{tabular} \\

\hline

BaTi$_{0.70}$Zr$_{0.30}$O$_3$ &
\begin{tabular}{c} 300 \end{tabular} &
\begin{tabular}{c} 3.553 \end{tabular} &
\begin{tabular}{c} 0.0068 \end{tabular} &
\begin{tabular}{c} 19 \end{tabular} &
\begin{tabular}{c} 4.16 \end{tabular} &
\begin{tabular}{c} 4.07 \end{tabular} &
\begin{tabular}{c} 4.60 \end{tabular} &
\begin{tabular}{c} 0.019 \end{tabular} &
\begin{tabular}{c} 3.1 \end{tabular} &
\begin{tabular}{c} 0.84 \end{tabular} \\

\hline

BaTi$_{0.75}$Zr$_{0.25}$O$_3$ &
\begin{tabular}{c} 300 \end{tabular} &
\begin{tabular}{c} 3.544 \end{tabular} &
\begin{tabular}{c} 0.0060 \end{tabular} &
\begin{tabular}{c} 20 \end{tabular} &
\begin{tabular}{c} 4.16 \end{tabular} &
\begin{tabular}{c} 4.07 \end{tabular} &
\begin{tabular}{c} 4.60 \end{tabular} &
\begin{tabular}{c} 0.015 \end{tabular} &
\begin{tabular}{c} 2.7 \end{tabular} &
\begin{tabular}{c} 0.49 \end{tabular} \\

\hline\hline

\end{tabular}
\end{table*}

\subsubsection{\label{results} Fit results}

For all the relaxor samples investigated, the refined values of
the buckling angle $\Theta_{Zr}$ range from 18 to 20$\rm^o$.
$\Theta_{Ti}$ being smaller (less than 10$\rm^o$), the ZrO$_6$
octahedra are therefore distorted differently depending on the
type of the third neighbors (Zr or Ti) of the Zr central atom.

In BTZ relaxors, the existence of a nearly constant buckling angle
whatever the Zr substitution rate $x$, together with the invariant
Zr-O1 distance (Sec. \ref{shell1}) result in a Zr-Zr distance of
the order of 4.15 \AA, which hardly varies with $x$. Of course,
the Zr-Zr distance is smaller in BTZ samples than in BaZrO$_3$,
since the oxygen atoms are aligned with the Zr-Zr bonds in the
latter compound. The Zr-Ti distance (4.07 $\rm\AA$) does not
change with $x$, and is found to be significantly smaller than the
Zr-Zr distance. The Zr-Ba distance, on its turn, increases
with~$x$. From the Zr-Ti distance and the buckling angle of the
Zr-O1-Ti bond, we can derive the Ti-O1 distance value at 300 K in
BTZ relaxors, which is of the order of 1.98~\AA. This corresponds
to the shortest Ti-O distance in the quadratic BaTiO$_3$ at 300 K
\cite{Kwei}.

\begin{figure}
\[\includegraphics[scale=0.4]{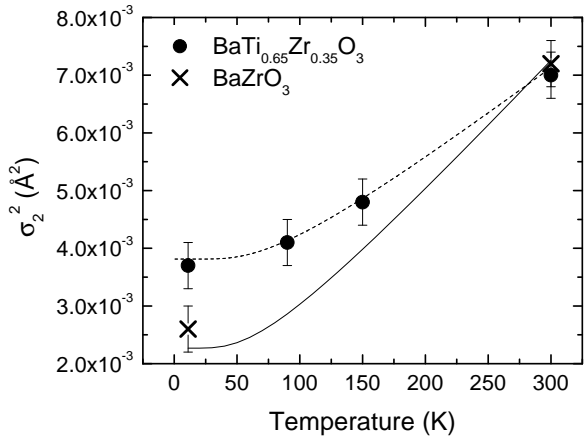}\]
\caption{Thermal evolution of the measured DW factor for the Zr-Ba
bond (symbols). The solid line represents Eq. \ref{equation2} with
$\theta_E$ = 194 K. The dashed line corresponds to the same
equation with $\theta_E$ = 231 K, and shifted by 0.0019 \AA$^2$.}
\label{sigma2}
\end{figure}

We now consider the temperature dependence of ${\sigma_{2}}^2$ in
BaTi$_{0.65}$Zr$_{0.35}$O$_3$ sample. As shown on Fig.
\ref{sigma2}, it can be described following Eq. \ref{equation2}
with an Einstein temperature equal to 231 $\pm$ 4 K, instead of
194 $\pm$ 8 K in BaZrO$_3$. This result indicates an increase of
Zr-Ba bond strength with increasing Ti-content. In addition, a
constant $\Delta{\sigma_2}^2$ = 0.0019 \AA$^2$ must be added to
Eq. \ref{equation2}, revealing the presence of static disorder.
The full width at half maximum of the corresponding gaussian
distance distribution would be equal to
$2\sqrt{2\ln2\Delta{\sigma_2}^2}$ = 0.10 \AA. However, let us
remind that other distance distributions can be expected (see Sec.
\ref{shell1}). The possible distribution of the Zr-Zr and Zr-Ti
bond lengths is out of the scope of this study, since
${\sigma_{3}}^2$ and ${\sigma_{3'}}^2$ parameters had to be fixed
during the refinement process.

The mean number of Zr third neighbors around a Zr atom $N_{Zr}$
brings information on the distribution of the Zr and Ti atoms in
BTZ relaxor samples. $N_{Zr}$ is found to be equal to 2.7, 3.1,
and 3.4 in BaTi$_{0.75}$Zr$_{0.25}$O$_3$,
BaTi$_{0.70}$Zr$_{0.30}$O$_3$, and BaTi$_{0.65}$Zr$_{0.35}$O$_3$
samples respectively. Despite the large uncertainty attached to
$N_{Zr}$ ($\pm$ 1.0), the measured values are higher than those
expected for a random distribution of Zr and Ti atoms: 1.5, 1.8,
and 2.1. This observation indicates a tendency of Zr atoms to
segregate in Zr-rich regions. However, the EXAFS technique giving
only an average of the local structure, we cannot conclude on the
Zr concentration in the Zr-rich regions, and hence on the sizes of
these regions.

In summary, in all the BTZ relaxors investigated the shape of a
ZrO$_6$ octahedra depends on the nature of its neighboring
octahedra (ZrO$_6$ or TiO$_6$), the oxygen atoms being differently
off-centered with respect to the Zr-Zr and Zr-Ti bonds. The Zr-Zr
and Zr-Ti distances do not depend on $x$, Zr-Ti distances being
significantly shorter than Zr-Zr ones. The Zr-Ba bond length
increases with the Zr-content, and is affected by a strong static
disorder. Finally, the EXAFS analysis reveals a tendency of Zr
atoms to segregate. Note that in the case of
BaTi$_{0.65}$Zr$_{0.35}$O$_3$, the Zr environment is found to be
remarkably stable with temperature. Therefore, no important
changes of the local structure are observed around Zr atoms in the
temperature range of the maximum of the dielectric permittivity.

\section{\label{disscu} Concluding discussion}

\subsection{\label{disscu1} Microstructural picture of BTZ relaxors}

\begin{figure}
\[\includegraphics[scale=0.4]{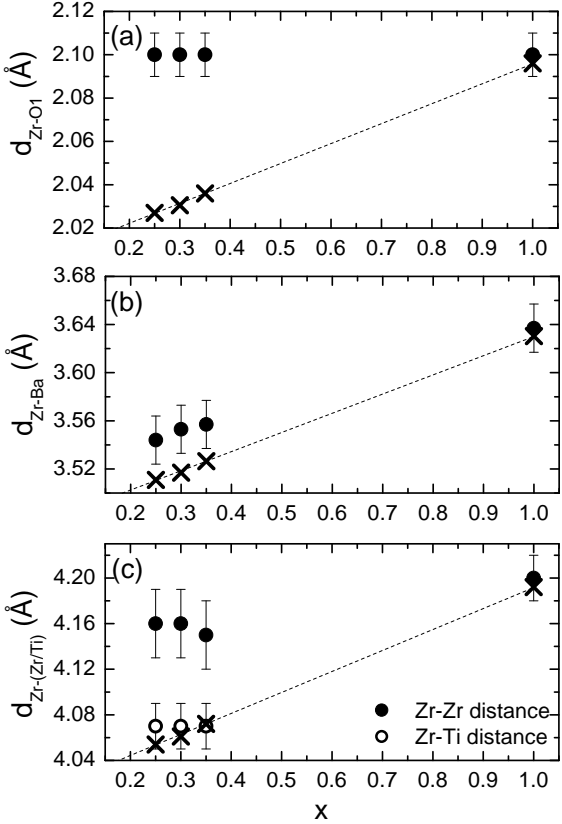}\]
\caption{Evolution of the Zr-O1 (a), Zr-Ba (b) and Zr-(Zr/Ti) (c)
distances with the Zr substitution rate $x$. EXAFS results are
reported as round dots. They are compared to the corresponding
average distances deduced from the X-ray diffraction experiments
(crosses); i.e. $a/2$, $a\sqrt{3}/2$, and $a$ respectively, where
$a$ is the cubic cell parameter.} \label{DistEXAFS-DRX}
\end{figure}

In the preceding sections we have described quantitatively the
local structural features of the Zr environment in BTZ relaxors.
It is now interesting to discuss how this local structure
contrasts with the average, long-range cubic structure evidenced
by X-ray diffraction (XRD). The interatomic distances deduced from
EXAFS are compared to those determined from XRD on Fig.
\ref{DistEXAFS-DRX}. The two techniques yield quite different
values of distances. On one hand, the XRD measurements show that
the unit cell dimensions of BTZ samples follow the Vegard's law,
i.e., the cell volume linearly increases from its value in
BaTiO$_3$ (64.286 $\rm\AA^3$) to that in BaZrO$_3$ (73.665
$\rm\AA^3$). In the cubic relaxor samples, all the average
distances then linearly increase with the Zr substitution rate
$x$. On the other hand, the distances deduced from the EXAFS
analysis systematically exceed those expected in the average
structure and, with the exception of the Zr-Ba distance, are found
to be independent of $x$. Considering that EXAFS probes the
structure on a very local scale, the latter observation is direct
evidence that the local structure is different from the average
structure.

We have shown in Sec. \ref{results} that Zr-atoms tend to
segregate in BTZ relaxors. The latter result is in agreement with
previous speculations based on Raman scattering \cite{RamanJens}.
If we assume that the Zr-rich regions consist of BaZrO$_3$
spherical inclusions in a BaTiO$_3$ matrix, one can calculate the
mean number of segregated atoms. It is estimated to 27, 15, and 10
in BaTi$_{0.65}$Zr$_{0.35}$O$_3$, BaTi$_{0.70}$Zr$_{0.30}$O$_3$,
and BaTi$_{0.75}$Zr$_{0.25}$O$_3$ respectively, which corresponds
to diameters of 14.1, 11.5, and 10.8~$\rm\AA$.

The EXAFS study of BTZ relaxors at the Zr K-edge allowed us to
evidence local structural deviations from the average cubic
structure, as well as chemical inhomogeneities. In the following,
we shall propose a microstructural picture of BTZ relaxors and
discuss its origin and implications on the relaxor behaviour. In
order to expose the possible basic physical mechanisms in a clear
and simple manner, we will assume that the segregation of Zr atoms
consists of small inclusions of BaZrO3 in a BaTiO3 matrix.
However, the real repartition of the Zr atoms could be much more
complex (see Sec. \ref{results}).

Such BaZrO$_3$ inclusions in BTZ relaxors must be submitted to a
chemical pressure from the BaTiO$_3$ environment, which has a
smaller unit cell volume than BaZrO$_3$. It is first interesting
to note that the Zr-O distance in BTZ relaxors is almost
insensitive to this chemical pressure, since it is the same as in
BaZrO$_3$ (Fig. \ref{DistEXAFS-DRX}-a). On the other hand, the
Zr-Ba distance decreases with increasing Ti-content (Fig.
\ref{DistEXAFS-DRX}-b) and hence, is more sensitive to the
chemical pressure. This latter difference in {\it AB}O$_3$
perovskites is well understood within the polyhedral approach: the
{\it A}-O distance in perovskites is generally more compressible
than the shorter {\it B}-O distance in the closed-packed BO$_6$
octahedra \cite{Hazen}. Insight into the {\it B}O$_6$ linkage can
be also gained from the $\rm \widehat{Zr O Zr}$ angle value, which
deviates considerably from 180$\rm^o$ (Table \ref{fitBTZ}). In a
perovskite structure, the presence of buckled Zr-O-Zr bonds
suggests the presence of octahedron tilts. This distortion could
be a direct consequence of the chemical pressure exerted by the
surrounding BaTiO$_3$ matrix. Within this context, let us remind
that hydrostatic pressure reveals and/or enhances structural
lattice instabilities related to tilts in the perovskite compounds
with {\it B}O$_6$ polyhedra more rigid than {\it A}O$_{12}$ ones
\cite{tilts}. Furthermore, it has been proposed in the literature
that BaZrO$_3$ undergoes a pressure-induced phase transition to
the tilted $a^-a^-a^-$ perovskite structure \cite{Lucazeau}
(Glazer's notations \cite{Glazer}). Assuming that the BaZrO$_3$
inclusions adopt the latter structure in BTZ relaxors, the tilt
angle calculated from the buckling angle of the Zr-O-Zr bonds is
about 11$\rm^o$, which is a reasonable value for a tilted
perovskite \cite{Tilt1,Tilt2,Tilt3}. In summary, within our
simplified picture, BaZrO$_3$ inclusions do not adopt the cubic
bulk structure of BaZrO$_3$ but present considerable distortions
away from it. The Zr environment in BTZ relaxors appears to be
determined by both the high stiffness of the ZrO$_6$ octahedra and
chemical pressure effects.

Let us now discuss the structure around the inclusions. Given that
the Zr-Zr distance is approximately 2\% larger than the average
{\it B}-{\it B} cation distance deduced from XRD (Fig.
\ref{DistEXAFS-DRX}-c), we expect a significant strain at the
interface of the BaZrO$_3$ inclusions and the BaTiO$_3$ matrix.
The TiO$_6$ octahedra adjacent to the inclusions likely
accommodate the structural misfit between BaZrO$_3$ inclusions and
the BaTiO$_3$ matrix. Considering that the cell volume of
BaZrO$_3$ is larger than that of BaTiO$_3$, we expect that these
adjacent TiO$_6$ octahedra are submitted to a tensile strain.
Although their deformation shape remains an open question,
different Ti$^{4+}$ displacements (and thus polarity) are expected
in the adjacent regions when compared to the BaTiO$_3$ matrix.

\subsection{\label{disscu} Comparison to PZT}

We have already mentioned in the introduction that the
ferroelectric PbZr$_{1-x}$Ti$_x$O$_3$ (PZT), similar to BTZ, does
surprisingly present no relaxation, making the comparison of BTZ
to PZT insightful. First, we note that in both BTZ and PZT, the
mean Zr-O distance is almost unaffected by the Ti-content, and the
observed Zr-O distance distribution is of the same order of
magnitude \cite{Cao}. As a consequence, the relaxor properties of
BTZ cannot be linked to the first neighbour environment of Zr
atoms.

A key feature for understanding the different behaviors of BTZ and
PZT could lie in the local chemical order, which is known to play
a determinant role for relaxor properties (e.g. the so-called 1:1
Mg/Nb order on the nanometer scale in PMN) \cite{Cross1,Cross2}.
In that context, it is interesting to note that BTZ cannot be
synthesized in the region $0.5 < x < 1$ (Ref. \onlinecite{Annie}),
which suggests an important internal chemical strain which
inhibits the formation of BTZ across the whole composition range.
We speculate that the local phase segregation observed for BTZ
relaxors is a precursor for the macroscopic phase separation
observed for higher Zr substitution rates. Such an internal
chemical strain does not seem to exist for PZT since it forms a
solid solution whatever the Zr/Ti ratio. The distribution of
elastic fields is thus expected to differ significantly from BTZ
to PZT. In particular, the strain around the local phase
separation in BTZ can be regarded as the source of the elastic
random fields which have been proposed earlier by Farhi et al
\cite{REF}. Such elastic random fields (similarly to electric
random fields in common relaxors) are then considered to lead to
particular local pattern of polar regions, which become the source
of relaxor properties.

\begin{acknowledgments}

We are indebted to O. Proux (LGIT) for his help during the EXAFS
experiments. This work is supported by an ACI project of the
french ministry of research, and has been conducted within the
framework of the European network of excellence FAME.

\end{acknowledgments}


\begin{thebibliography}{99}

% Introduction

\bibitem{SingleCryst} P. W. Rehrig, S.E. Park, S. Trollier-McKinstry,
G. L. Messing, B. Jones, T. R. Shrout, \JAP {\bf 86}, 1657 (1999).
\bibitem {BTZpiezoOrient} Z. Yu, R. Guo, A. S. Bhalla, Appl. \PL {\bf
77}(10), 1535 (2000).
\bibitem {Poudre} Z. Yu, C. Ang, R. Guo, A.S. Bhalla, \JAP {\bf 92},
1489 (2002).
\bibitem {DRAM1} T.B. Wu, C.M. Wu, M.L. Chen, Thin Solid Films {\bf 334}, 77 (2003).
\bibitem {DRAM2} V. Reymond, S. Payan, D. Michau, J.P. Manaud, M. Maglione, Thin Solid Films {\bf 467}, 54 (2004).
\bibitem {BTZorder} Y. Hotta, G. W. J. Hassink, T. Kawai, H. Tabata,
Jap. \JAP {\bf 42}(9B), 5908 (2003).
\bibitem {Annie} J. Ravez , A. Simon, Eur. J. Solid State Inorg. Chem.
{\bf 34}, 1199 (1997).
\bibitem {Park} S. E. Park, T. R. Shrout, \JAP {\bf 82}, 1804 (1997).
\bibitem {LeadfreeGiant} Y.-M. Chiang, G.W. Farrey, A. N. Soukhojak,
Appl. \PL {\bf 73}(25), 3683 (1998).
%\bibitem {Smolenskii} G. A. Smolenskii, A. I. Agranovskaya, Soviet Physics Solid State {\bf 1}, 1429 (1959).
\bibitem {Cross1} L. E. Cross, Ferroelectrics {\bf 76}, 241 (1987).
\bibitem {Cross2} L. E. Cross, Ferroelectrics {\bf 151}, 305 (1994).
\bibitem{Samara} G. A. Samara, \JPCM {\bf 15}, R367 (2003).
\bibitem{Westphal} V. Westphal, W. Kleemann, M. D. Glinchuk, \PRL {\bf 68}, 847 (1992).
\bibitem{Burns} G. Burns, F. H. Dacol, \SSC {\bf 48}, 853 (1983).
\bibitem {Bonneau} P. Bonneau, P. Garnier, G. Calvarin, E. Husson, J.
R. Gavarri, A. W. Hewat, A. Morell, Journal of Solid State
Chemistry {\bf 91}, 350 (1991).
\bibitem {deMathan} N. de Mathan, E. Husson, G. Calvarin, J. R.
Gavarri, \JPCM {\bf 3}(42), 8159 (1991).
\bibitem {PMN_RS} S. Vakhrushev, A. Nabereznov, S. K. Sinha, Y. P. Feng, T. Egami, \JPhCh Solids {\bf 57}, 1517 (1996).
\bibitem {PMN_Diff_HP_Jens} B. Chaabane, J. Kreisel, B. Dkhil, P. Bouvier, M. Mezouar, \PRL {\bf 90}, 257601 (2003).
\bibitem {PMN_Diff_neutrons} G. Xu, G. Shirane, J. R. D. Copley, P. M. Gehring, {\PR} B {\bf 69}, 064112 (2004).
\bibitem {PMN_PDF} I. K. Jeong, T. W. Darling, J. K. Lee, T. Proffen, R. H. Heffner, J. S. Park, K. S. Hong, W. Dmowski, T. Egami, \PRL {\bf 94}, 147602 (2005).
\bibitem {REF} R. Farhi, M. El Marssi, A. Simon, J. Ravez, {\EPJ} B {\bf 18}, 605 (2000).
\bibitem {Farhi} R. Farhi, M. El Marssi, A. Simon, J.Ravez, {\EPJ} B {\bf 9}, 599 (1999).
\bibitem {RamanJens} J. Kreisel, P. Bouvier, M. Maglione, B. Dkhil, A. Simon, {\PR} B {\bf 69}, 092104 (2004).
\bibitem {Sciau} P. Sciau, G. Calvarin, J. Ravez, Solid State
Communications {\bf 113}, 77 (2000).
\bibitem {BTZtrans} A. Simon, J. Ravez, M. Maglione, \JPCM {\bf 16},
963 (2004).
\bibitem {BTZpression} J. Kreisel, P. Bouvier, M. Maglione, B. Dkhil,
A. Simon, {\PR} B {\bf 69}, 092104 (2004).
\bibitem {Noheda} B. Noheda, Current Opinion in Solid State and
Materials Science {\bf 6}, 27 (2002).
\bibitem {KTN_XAFS_Temp} O. Hanske-Petitpierre, Y. Yacoby, J. Mustre de Leon, E. A. Stern, J. J. Rehr, {\PR} B {\bf 44}, 6700 (1991).
\bibitem {KNO_XAFS_P} A. I. Frenkel, F. M. Wang, S. Kelly, R. Ingalls, D. Haskel, E. A. Stern, Y. Yacobi {\PR} B {\bf 56}, 10869 (1997).
\bibitem {PTO_XAFS_Temp} N. Sicron, B. Ravel, Y. Yacoby, E. A. Stern, F. Dogan, J. J. Rehr, {\PR} B {\bf 50}, 13168 (1994).
\bibitem{Cao} D. Cao, I. -K. Jeong, R. H. Heffner, T. Darling, J. -K. Lee, F. Bridges, J. -S. Park, K. -S. Hong, {\PR} B {\bf 70}, 224102 (2004).
\bibitem {PMN-XAFS} E. Prouzet, E. Husson, N. de Mathan, A. Morell,
\JPCM {\bf 5}, 4889 (1993).
\bibitem {Relaxor-XAFS} I. W. Chen, P. Li, Y. Wang, \JPhCh in Solids
{\bf 57}(10), 1525 (1996).
\bibitem{XAFSmixed} V. A. Shuvaeva, I. Pirog, Y. Azuma, K. Yagi, K.
Sakaue, H. Terauchi, I.P. Raevski, K. Zhuchkov, M. Y. Antipin,
\JPCM {\bf 15}, 2413 (2003).
\bibitem{XAFSPIN} V. A. Shuvaeva, Y. Azuma, I.P. Raevski, K. Yagi, K.
Sakaue, H. Terauchi, Ferroelectrics {\bf 299}, 103 (2004).
\bibitem {KNBT-XAFS} V. A. Shuvaeva, D. Zekria, A. M. Glazer, Q. Jiang,
S. M. Weber, P. Bhattacharya, P. A. Thomas, {\PR} B {\bf 71},
174114 (2004).

%Partie II B

\bibitem {Autobk} M. Newville, P. Livins, Y. Yacoby, J. J.
Rehr, E. A. Stern, {\PR} B {\bf 47}, 14126 (1993).
\bibitem {Zabinsky} S. I. Zabinsky, J. J. Rehr,A. Ankudinov, R. C.
Albers, M. J. Eller, {\PR} B {\bf 52}, 2995 (1995).
\bibitem {FEFFIT} M. Newville, B. Ravel, D. Haskel, J. J. Rehr, E. A.
Stern, and Y. Yacoby, Physica B {\bf 208-209}, 154 (1995).
\bibitem {FEFF} A.L. Ankudinov, B. Ravel, J.J. Rehr, and S.D.
Conradson, Phys. Rev. B {\bf 58}, 7565 (1998).

%Partie III A

\bibitem {Ravel} D. Haskel, B. Ravel, M. Newville, E. A. Stern, Physica
B {\bf 208-209}, 151 (1995).
\bibitem {Bugaev} L. A. Bugaev, V. A. Shuvaeva, I. B. Alekseenko, R. V.
Vedrinskii, Physica B {\bf 208-209}, 169 (1995).
\bibitem {Petit} P. E. Petit, F. Guyot, F. Farges, J. Phys. IV France
{\bf 7}, C2-1065 (1997).
\bibitem {footnote1} Actually, small local deviations from a perfect cubic
structure do exist in BaZrO$_3$, as shown by the existence of a
weak first-order Raman spectrum (see C. Chemarin and al, J. Sol.
State Chem., {\bf 149}, 298 (2000)). Our purpose being to set
reference parameters for the analysis of BTZ, these minor
distortions can be safely neglected here.
\bibitem{Mathews} M. D. Mathews, E. B. Mirza, A. C. Momin, Journal of
Material Science Letters {\bf 10}, 305 (1991).
\bibitem{Ba-CaZrO3struct} I. Levin, T. G. Amos, S. M. Bell, L. Farber,
T. A. Vanderah, R. S. Roth, B. H. Toby, Journal of Solid State
Chemistry {\bf 175}, 170 (2003).

%Partie III C 1

\bibitem{footnote2} Different energy shifts ${\Delta E_0}_i$ could have been introduced for
each of the SS paths, in order to compensate for slight
inadequancies of the calculated phase shifts \cite{Ravel}. We do
not use this method, our aim being to compare the distances
obtained in BaZrO$_3$ and BTZ samples. In the case of BTZ relaxors
indeed, the complexity of the fit is such that these additional
parameters cannot be meaningful.

%Partie III C 2

\bibitem{Frenkel} A. Frenkel, E.A. Stern, A. Voronel, M. Qian, M. Newville, {\PR} B {\bf 49}, 11662 (1994).
\bibitem{Kwei} G.H. Kwei, A.C. Lawson, S.J.L. Billinge, S.W. Cheong, J. Phys. Chem. {\bf 97}, 2368 (1993).

%Discussion

\bibitem{Hazen} Comparative Crystal Chemistry, R.M. Hazen and L.W. Finger, Wiley (New york, 1982).
\bibitem{tilts} R.J. Angel, J. Zhao, N.L. Ross, \PRL {\bf 95}, 025503 (2005).
\bibitem {Glazer} A. M. Glazer, \AC B {\bf 28}, 3384 (1972).
\bibitem{Lucazeau} C. Chemarin, N. Rosman, T.Pagnier, G. Lucazeau, J. Sol. State Chem. {\bf 149}, 298 (2000).
\bibitem{Tilt1} Perovskites - Modern and Ancient, R.H. Mitchell, Almaz Press (Thunder Bay, 2002).
\bibitem{Tilt2} K.A. M\"{u}ller, W. Berlinger, F. Waldner, \PRL {\bf 21}, 814 (1968).
\bibitem{Tilt3} G. Roult, R. Pastuszak, R. Marchand, Y. Laurent, Acta Cryst. C {\bf 39}, 673 (1983).

\end{thebibliography}
\end{document}